\documentclass[australian,english,prl, singlespace, twocolumn]{revtex4-1}
\usepackage[T1]{fontenc}
\usepackage[latin9]{inputenc}
\usepackage{color}
\usepackage{amsmath}
\usepackage{amssymb}
\usepackage{graphicx}

\makeatletter
\@ifundefined{definecolor}
 {\usepackage{color}}{}
\makeatother

\makeatother

\usepackage{babel}
\begin{document}
\title{Quantifying the mesoscopic nature of the Einstein-Podolsky-Rosen nonlocality}
\author{M. D. Reid$,^{1,2}$}
\email{mdreid@swin.edu.au}

\author{Q. Y. He$^{3,4}$}
\email{qiongyihe@pku.edu.cn}

\affiliation{$^{1}$Centre for Quantum and Optical Science Swinburne University
of Technology, Melbourne, Australia}
\affiliation{$^{2}$Institute of Theoretical Atomic, Molecular and Optical Physics
(ITAMP),Harvard University, Cambridge, Massachusetts, USA}
\affiliation{$^{3}$State Key Laboratory of Mesoscopic Physics, School of Physics,
Peking University, Collaborative Innovation Center of Quantum Matter,
Beijing 100871, China}
\affiliation{$^{4}$Collaborative Innovation Center of Extreme Optics, Shanxi University,
Taiyuan 030006, China}
\begin{abstract}
Evidence for Bell's nonlocality is so far mainly restricted to microscopic
systems, where the elements of reality that are negated predetermine
results of measurements to within one spin unit. Any observed nonlocal
effect (or lack of classical predetermination) is  then limited to
no more than the difference of a single photon or electron being detected
or not (at a given detector). In this paper, we analyze experiments
that report Einstein-Podolsky-Rosen (EPR) steering form of nonlocality
for mesoscopic photonic or Bose-Einstein condensate (BEC) systems.
Using an EPR steering parameter, we show how the EPR nonlocalities
involved can be quantified for four-mode states, to give evidence
of nonlocal effects corresponding to a two-mode number difference
of $10^{5}$ photons, or of several tens of atoms (at a given site).
We also show how the variance criterion of Duan-Giedke-Cirac and
Zoller for EPR entanglement can be used to determine a lower bound
on the number of particles in a pure two-mode EPR entangled or steerable
state, and apply to experiments.
\end{abstract}
\maketitle
Einstein-Podolsky-Rosen (EPR) presented in 1935 a seemingly compelling
argument that quantum mechanics was incomplete \cite{epr}.  In their
gedanken experiment, properties of a system $B$ can be predicted
ultra-precisely, by the measurements of a distant observer, popularly
called Alice. EPR assumed no ``spooky action-at-a-distance'', to
argue that Alice's measurement is noninvasive, and therefiore that
Alice's prediction represents a predetermined property (an ``element
of reality'') of system $B$. Further, they showed that the set of
all such predetermined properties could not be consistent with any
local quantum state description for $B$, and thus concluded that
quantum mechanics was an incomplete theory. The assumptions made in
EPR's argument are collectively known as local realism (LR). Bell's
theorem negated these premises, by showing  LR could be falsified
\cite{Bell}.

\textcolor{black}{Understanding whether and how local realism fails
macroscopically remains an open question in physics. Loophole-free
experiments confirming Bell's theorem are so far limited to microscopic
systems e.g. system $B$ is a single photon or electron \cite{bell-ep,loophole}.
In these cases, the predetermined properties that EPR called elements
of reality give predictions to within a single spin unit. The failure
of LR that is inferred from the experiments is therefore a microscopic
effect only, in the sense that this  pertains  only to predictions
specified to an accuracy of one spin unit, for a microscopic particle.
Similar accuracies are required  in almost all of the experiments
predicted to violate LR for multi-particle systems \cite{higher-n-bell,cv-bell}.}

\textcolor{black}{By contrast, EPR's experiment (called an ``EPR-steering''
experiment \cite{schosteer,hw,steereric,eprr,saunexp}) has been investigated
experimentally for mesoscopic optical fields \cite{ou,rrmp,spinpolepr,Aiko_2010_PRA_exp(aa),Tobias2011_cc,Walborn2011,Aoki_2010_(a),Hannover2012best_(d),Hannover2013best,Howell2013,pengbest_(b),recentD_(c),photon-pair-2016},
atomic ensembles \cite{eprnaturecommun,SteerAt-obert,treu-matteo,Plozik,Polzik2011EPR_macro_(e),muschik,muschik2,Optica2017},
and, recently, for mesoscopic mechanical oscillators \cite{paulo,eprentosc,polamaki,roman-schnabel-ent-epr-objects,recent-ent-osc,recnet-osci-2,HeReid2013,Simon2014,bell-osc}.
In many of these experiments, not only are the systems sizeable, but
the outcomes are over a larger range, corresponding to several or
many spin units. There is thus the possibility to test for a mesoscopic
EPR nonlocal effect, where the predetermined ``elements of reality''
that are falsified give predictions with an indeterminacy of several
spin units. One may then ask how much ``spooky action-at-a-distance''
is occurring in terms of spin units? How to do the quantification
is not obvious, however. It is not simply the size of the entangled
system, nor the range of outcomes. Previous measures inform us how
many atoms are mutually entangled \cite{sorsolm-1,multi}, or what
fraction of particle pairs behave locally versus nonlocally \cite{other-measures},
but these need not imply large differences in the actual }\textcolor{black}{\emph{outcomes}}\textcolor{black}{{}
of observables due to nonlocal effects.}\textcolor{red}{}

The situation is clear if the physical quantities measured by observers
Bob and Alice (at different location\textcolor{black}{s) have two
mesoscopically-distinct outcomes $+$ and $-$, e.g. $N$ particles
in an }\textcolor{black}{\emph{up}}\textcolor{black}{{} position versus
$N$ particles in a }\textcolor{black}{\emph{down}}\textcolor{black}{{}
position ($N\gg1$) \cite{Schrodinger,LG}. One may then extend EPR's
premises, to define $\delta$-scopic local realism ($\delta$LR),
which asserts \cite{mlr,recent-mlr}: (1) any measurement by Alice
cannot instantly induce a change $\delta=2N$ to the outcome of measurement
at Bob's location; and (2) if the outcome $+$ or $-$ for Bob's system
can be predicted with certainty by Alice, then Bob's system is always
predetermined to be in a state that gives }\textcolor{black}{\emph{either}}\textcolor{black}{{}
the result $+$ or $-$. The failure of such premises then implies
a ``spooky action'' effect of size $\delta=2N$.  However, examples
of EPR systems with just two outcomes $\pm$ separated by $\delta$
are limited.}

\textcolor{black}{In this paper, we present a practical approach to
test EPR premises based on $\delta$-scopic local realism ($\delta$LR).
We consider a version of EPR's argument based on the premise of $\delta$LR,
where the separation of outcomes $+$ and $-$ is quantified by $\delta$,
but where there is a continuous range of outcomes. Our analysis leads
to a criterion sufficient to demonstrate EPR $\delta$-scopic nonlocality,
that there is inconsistency between the completeness of quantum mechanics
and $\delta$LR. The size of $\delta$  quantifies in the $\delta$LR
assumption the upper bound on the amount of change  that can occur
to Bob's system, due to Alice's measurements. Failure of $\delta$LR
where $\delta$ is large implies large nonlocal effects. We analyze
EPR experiments which have a mesoscopic, continuous range of outcomes
for Alice and Bob's measurements, to present preliminary evidence
for quantifiable mesoscopic EPR nonlocalities.}

\textbf{\textcolor{black}{Quantifying the EPR Paradox:}}\textcolor{black}{{}
The EPR argument can be generalised to pairs of measurements $\{X_{A},P_{A}\}$
and $\{X_{B},P_{B}\}$ on two spatially separated systems $A$ and
$B$. We consider $X_{B}$, $P_{B}$ to be scaled non-commuting observables
which satisfy $[X_{B},P_{B}]=2$, so that the Heisenberg uncertainty
relation is $\Delta X_{B}\Delta P_{B}\geq1$. To demonstrate the paradox,
one measures the variances $V_{B}(X_{B}|X_{A})$and $V_{B}(P_{B}|P_{A})$
of the respective conditional distributions $P(X_{B}|X_{A})$ and
$P(P_{B}|P_{A})$ \cite{eprr}. Here $P(X_{B}|X_{A})$ is the probability
for result $X_{B}$, given a measurement $X_{A}$. The average conditional
variance $(\Delta_{inf}X_{B})^{2}=\sum_{X_{A}}P(X_{A})V_{B}(X_{B}|X_{A})$
determines the accuracy of inference of the results for $X_{B}$,
based on the measurements at $A$. The $\Delta_{inf}P_{B}$ is defined
similarly. Using EPR's logic, these inference variances define the
average indeterminacy of the two respective elements of reality, for
$X_{B}$ and $P_{B}$. If 
\begin{equation}
\varepsilon\equiv\Delta_{inf}X_{B}\Delta_{inf}P_{B}<1\label{eq:epr-cond}
\end{equation}
then an EPR paradox arises, since the simultaneous predetermination
for $X_{B}$ and $P_{B}$ is more accurate than allowed by the uncertainty
principle \cite{rrmp,eprr}. The condition (1) is a condition for
``}\textcolor{black}{\emph{EPR steering}}\textcolor{black}{{} of system
$B$'' \cite{hw,steereric}.}

\textcolor{black}{We now construct a quantified version of the EPR
argument, by relaxing EPR's premises. The assumptions of $\delta_{X}$-scopic
local realism ($\delta_{X}$LR) are: (1) A measurement made at $A$
might disturb the system $B$, so that the outcome for a simultaneous
measurement $X$ on $B$ can be altered, but the change to the outcome
cannot be greater (in magnitude) than $\delta_{X}$. (2) If the value
for a physical quantity $X$ is predictable, without disturbing the
system by more than $\delta_{X}$, then the value of that physical
quantity is a predetermined property of the system (the ``element
of reality'' for $X$), the predetermined value being given to within
$\pm\delta_{X}$ of the predicted value.  We will refer to $\delta_{X}$
as the degree of ``}\textcolor{black}{\emph{nonlocal indeterminacy}}\textcolor{black}{'',
with respect to the EPR observable $X$. }

The assumption of $\delta$LR changes the condition for an EPR paradox,
making it more difficult to demonstrate the paradox. Applying $\delta_{X}$LR,
the indeterminacy in the predictions for $X_{B}$ associated with
the element of reality has increased, but by a limited amount only.
We show \textcolor{black}{in the Supplemental materials }that the
maximum value of this indeterminacy becomes \cite{supp} 
\begin{eqnarray}
(\Delta_{inf,\delta_{X}}X_{B})^{2} & = & (\Delta_{inf}X_{B})^{2}+\delta_{X}^{2}+\nonumber \\
 &  & 2\delta_{X}\sum_{X_{A},X_{B}}P(X_{A},X_{B})|X_{B}-\langle X_{B}|X_{A}\rangle|,\nonumber \\
\label{eq:neweprdeltaline}
\end{eqnarray}
where $P(X_{A},X_{B})$ is the joint probability. Defining $\Delta_{inf,\delta_{P}}P_{B}$
in a similar manner, the experimental realisation of
\begin{equation}
\varepsilon_{\delta}\equiv\Delta_{inf,\delta_{X}}X_{B}\Delta_{inf,\delta_{P}}P_{B}<1\label{eq:eprdelta-1}
\end{equation}
will therefore imply an inconsistency between the premise of $\delta$LR
and the com\textcolor{black}{pleteness of quantum mechanics. The
calculation of $\varepsilon_{\delta}$ is straightforward, once the
distributions $P(X_{A},X_{B})$ and $P(P_{A},P_{B})$ are known. When
$\delta=0$, Eq. (3) reduces to the standard EPR condition (1). The
inequality is progressively more difficult to satisfy, as $\delta$
increases.}

\textbf{\textit{Gaussian $\delta$-scopic EPR nonlocality:}} We
consider EPR experiments based on field modes at locations $A$, $B$.
$X_{A/B}$, $P_{A/B}$ are defined according to $a=(X_{A}+iP_{A})/2$
and $b=(X_{B}+iP_{B})/2$, where $a$, $b$ are the annihilation operators
of each mode.\textcolor{black}{{} The $\delta$-scopic EPR inequality
reduces to Eq. (\ref{eq:eprdelta-1}).} A widely-used source of EPR-correlated
fields is the parametric amplifier, the ideal output of which is the
two-mode squeezed state \cite{rrmp,eprr}. Here, the conditionals
$P(X_{B}|X_{A})$ and $P(P_{B}|P_{A})$ are Gaussian. Moreover, a
Gaussian profile is maintained in non-ideal situations where losses
and thermal noise are present \cite{rrmp,cv-review,gauscv}.

\textcolor{black}{Assuming Gaussianity, the prediction of $\varepsilon_{\delta}$
given measured values of} $\Delta_{inf}X_{B}$ and $\Delta_{inf}P_{B}$\textcolor{black}{{}
is straightforward. Using }Eq. (\ref{eq:neweprdeltaline})\textcolor{black}{{}
and that for a Gaussian distribution $\langle|X_{B}-\mu_{X}|\rangle=\Delta_{inf}X_{B}\sqrt{2/\pi}$}
(where $\mu_{X}$ is the mean of $P(X_{B}|X_{A})$), we find $\varepsilon_{\delta}=\sigma^{2}+\delta^{2}+2\delta\sigma\sqrt{2/\pi}$
\cite{supp}. For the sake of simplicity, we have taken $\sigma=\Delta_{inf}X_{B}=\Delta_{inf}P_{B}$
and $\delta=\delta_{X}=\delta_{P}$. We see that $\varepsilon{\color{black}<\left[-\delta\sqrt{2/\pi}+\sqrt{2\delta^{2}/\pi-(\delta^{2}-1)}\right]^{2}}$
will be \textcolor{black}{sufficient to imply }the $\delta$-scopic
EPR nonlocality.

\begin{figure}[h]
\begin{centering}
\includegraphics[width=1\columnwidth]{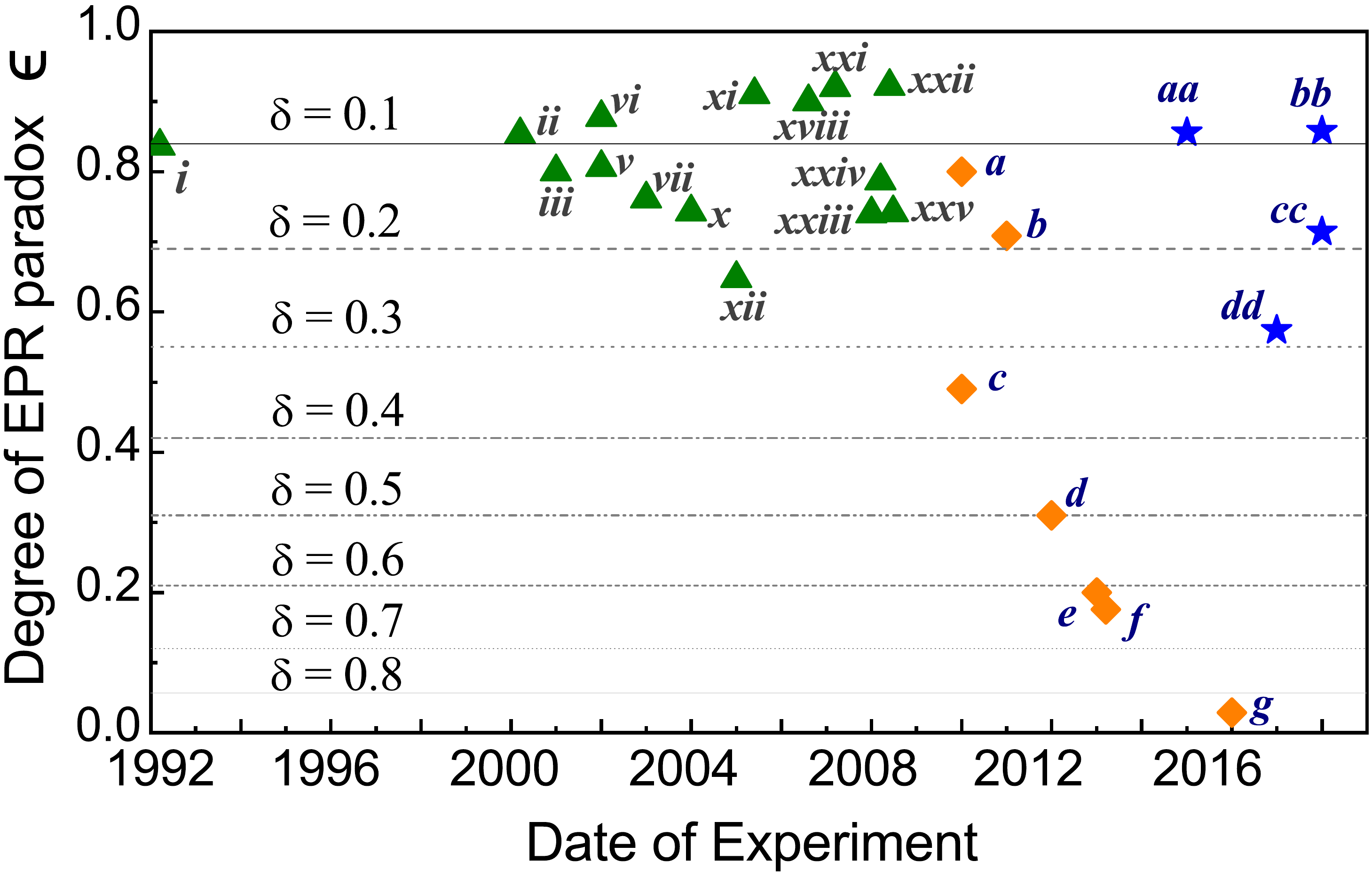}
\par\end{centering}
\caption{ The $\delta$-scopic EPR nonlocality is realised ($\epsilon_{\delta}<1$)
when $\epsilon$ is below the line shown, for the given $\delta$.
\textcolor{black}{Data $i-xxv$ are for the experiments referenced
in Fig. 9 of Ref. \cite{rrmp}, while data $a$ (\cite{Aiko_2010_PRA_exp(aa)}),
$b$ (\cite{Tobias2011_cc}), $c$ (\cite{pengbest_(b)}), $d$ (\cite{recentD_(c)}),
$e$ (\cite{Hannover2012best_(d)}), $f$ (\cite{Hannover2013best}),
$g$ (\cite{photon-pair-2016}), $aa$ (\cite{eprnaturecommun}),
$bb$ (\cite{SteerAt-obert}), $cc$ (\cite{treu-matteo}), $dd$
(\cite{Optica2017}) are later experiments. All results are for spatially
separated optical fields, except those given by blue stars which are
for mesoscopic groups of cooled atoms ($aa,$ $bb$, $cc$) or for
hybrid systems ($dd$). The cold atom groups have no (as in $aa$)
or small spatial separations of $\sim10$$\mu$m ($bb$, $cc$). \label{fig:Data-for-EPR}}}
\end{figure}

Extensive data has been reported for continuous variable EPR experiments
\textcolor{black}{\cite{ou,rrmp,spinpolepr,Aiko_2010_PRA_exp(aa),Tobias2011_cc,Walborn2011,Aoki_2010_(a),Hannover2012best_(d),Hannover2013best,Howell2013,eprnaturecommun,SteerAt-obert,treu-matteo,recentD_(c),pengbest_(b),photon-pair-2016}}
(see Fig. 1). Gaussian distributions are predicted in almost all
cases plotted \textcolor{black}{(including the data indicated by (g))}
as has been verified experimentally \cite{gauscv,rrmp}. For rigorous
testing, a full construction of the distributions with space-like
separated measurement events is required \cite{Bell,loophole}. \textcolor{black}{With
this proviso, we note that the recently achieved values of the EPR
parameter  $\varepsilon\sim0.176$ \cite{Hannover2013best} will
imply a $\delta$-scopic EPR nonlocality, with $\delta\sim0.633$.} 

To determine the\emph{ }significance of the value of $\delta$, one
needs to resort to the details of the individual experiments. The
nonlocal indeterminacy $\delta$ is given relative to the quantum
noise level, which for the optical experiments is usually considered
microscopic. On the other hand, entanglement has now been detected
 between two mechanical oscillators \textcolor{black}{\cite{paulo,eprentosc,recnet-osci-2},
}and between an oscillator and a field \cite{polamaki}. Entanglement
however does not imply the EPR steering condition (\ref{eq:epr-cond}).
It has been proposed to detect the EPR condition (\ref{eq:epr-cond})
for these cases \cite{paulo,roman-schnabel-ent-epr-objects,HeReid2013,Simon2014},
where $X_{B}$, $P_{B}$ refer to the quadratures of the \textcolor{black}{phonon
modes of the oscillator.}  Eq. (\ref{eq:eprdelta-1}) enables a quantification
of the EPR nonlocality that would be observed in such an experiment.
Here,\textcolor{red}{{} }\textcolor{black}{$\delta$ is quantifiable
at the Planck scale \cite{probeplanck}, and corresponds to a nonlocal
indeterminacy with respect to mechanical motion.}

\textbf{\emph{EPR nonlocality using Schwinger spins:}} For some experiments,
the quantum noise level and hence $\delta$ may correspond to a large
number of photons. This is understood by considering the Heisenberg
relation $\Delta J^{Z}\Delta J^{Y}\geq\bigl|\langle J^{X}\rangle\bigr|/2$
for spin systems, where measurements are made of the spin components
$J^{X,Y,Z}$. For high spins, $\bigl|\langle J^{X}\rangle\bigr|$
can be large. 

Indeed, EPR states exist for which $\bigl|\langle J^{X}\rangle\bigr|$
is a scalable large number. In these cases, the EPR observables are
two-mode Schwinger spins, defined as $J_{A}^{X}=(a_{+}^{\dagger}a_{-}+a_{+}a_{-}^{\dagger})/2$,
$J_{A}^{Y}=(a_{+}^{\dagger}a_{-}-a_{+}a_{-}^{\dagger})/2i$, $J_{A}^{Z}=(a_{+}^{\dagger}a_{+}-a_{-}^{\dagger}a_{-})/2$,
and $J_{B}^{X}=(b_{+}^{\dagger}b_{-}+b_{+}b_{-}^{\dagger})/2$, $J_{B}^{Y}=(b_{+}^{\dagger}b_{-}-b_{+}b_{-}^{\dagger})/2i$,
$J_{B}^{Z}=(b_{+}^{\dagger}b_{+}-b_{-}^{\dagger}b_{-})/2$, where
$a_{\pm}$, $b_{\pm}$ are annihilation operators for four modes \cite{mlr}.
The four modes are created from spatially separated modes $a$, $b$
prepared in an EPR state $|\psi\rangle$. Each mode $a$ $b$ interferes
(via a beam splitter) with an intense ``local oscillator'' field
(denoted by annihilation mode operators $b_{LO}$, $a_{LO}$). This
creates a macroscopic photonic state $|\psi\rangle_{M}$ involving
four fields $a_{\pm}=(a\pm a_{LO})/\sqrt{2}$, $b_{\pm}=(b\pm b_{LO})/\sqrt{2}$
at sites $A$ and $B$ respectively. The fields at each site pass
through second polarising beam splitters set at respective angles
$\theta_{A}$ and $\theta_{B}$. The number of particles in each arm
is detected, as a large number, and the difference gives a measure
of $J^{Z}$, $J^{Y}$ or $J^{Z}$ depending on the choice of $\theta_{A},\theta_{B}$.
Based on the Heisenberg uncertainty relation, the EPR criterion is
\begin{equation}
\Delta_{inf,\delta_{J}}(J_{B}^{Z})\Delta_{inf,\delta_{J}}(J_{B}^{Y})<|\langle J_{B}^{X}\rangle|/2,\label{eq:schepr}
\end{equation}
which normalises to Eq. (\ref{eq:eprdelta-1}) on defining $X_{B}/P_{B}=J_{B}^{Z/Y}/\sqrt{|\langle J_{B}^{X}\rangle|/2}$
and $\delta=\delta_{J}/\sqrt{|\langle J_{B}^{X}\rangle|/2}$. Here,
$|\langle J_{B}^{X}\rangle|=|\langle b_{LO}^{\dagger}b_{LO}-b^{\dagger}b\rangle/2|$
which becomes $\langle b_{LO}^{\dagger}b_{LO}\rangle/2$ since $\langle b^{\dagger}b\rangle/\langle b_{LO}^{\dagger}b_{LO}\rangle$
is small. The intensity of the local oscillator is macroscopic and
$\delta_{J}\sim\delta\sqrt{\langle b_{LO}^{\dagger}b_{LO}\rangle/4}$
(the nonlocal indeterminacy in the values of $J_{X}^{B}$, $J_{Y}^{B}$)
can therefore also be large.

EPR nonlocality for Schwinger spins has been realised in the experiments
of Bowen et al \cite{spinpolepr}, where $a_{\pm}$ ($b_{\pm}$) correspond
to two orthogonal horizontally (``H'' or ``$x$'') and vertically
(``V'' or ``$y$'') polarised field modes at $A$ ($B$). From
the description of their experiment \cite{spinpolepr,supp}, $|\langle J_{B}^{X}\rangle|\sim10^{11}$
photons, implying a $\delta_{J}$ of order $10^{5}$ photons. The
relative value $\delta_{J}/|\langle J_{B}^{X}\rangle|$ is however
small.

\textbf{\emph{Analogy to Schr}}\textbf{\textit{ö}}\textbf{\emph{dinger
cat:}} The Schwinger-spin experiment provides a simple parallel to
Schrödinger's cat gedanken experiment \cite{Schrodinger,mlr}. In
the original cat paradox, a macroscopic superposition is created by
the process of measurement, which couples the microscopic system (prepared
in a superposition state) to a measurement apparatus. In the experiment,
the microscopic EPR state $|\psi\rangle$ is indeed coupled to a macroscopic
system (the l\textcolor{black}{ocal oscillator fields) at each site,
and a four-mode amplified state $|\psi\rangle_{M}$ is produced that
enables a macroscopic readout of $X_{A/B}$ and $P_{A/B}$ of the
original fields.}

\textcolor{black}{The many-particle state $|\psi\rangle_{M}$ is created
prior to the measurements $J_{A}^{\theta_{A}}$, $J_{B}^{\theta_{B}}$
and it is this feature that enables the demonstration of mesoscopic
nonlocality. The $|\psi\rangle_{M}$ is a superposition of states
with definite outcomes for $J_{B}^{\theta_{B}}$, where those outcomes
are given by $J_{B}^{\theta_{B}}=EX_{B}/2$, or $EP_{B}/2$ (here
$E^{2}=\langle b_{LO}^{\dagger}b_{LO}\rangle$) depending on $\theta_{B}$.
The superposition $|\psi\rangle_{M}$ comprises many states that have
a large range of continuous outcomes for $J_{B}^{\theta_{B}}$, rather
than just two distinct states as in Schrodinger's case. In the Supplemental
Materials we prove that the observation of the $\delta_{J}$-scopic
EPR nonlocality can only arise if $|\psi\rangle_{M}$ comprises at
least two states that differ in outcome for $J_{B}^{\theta_{B}}$
by at least $\delta_{J}$ \cite{supp}. Such states (when $\delta_{J}$
is large) give a nonzero mesoscopic quantum coherence, and signify
a generalised Schrodinger-cat paradox (refer Refs. \cite{cat-sup-in-cv,cavalr,frowis,flor-cat-rmp}).
These states, for which EPR nonlocality is also demonstrated, are
well nested }\textcolor{black}{\emph{within}}\textcolor{black}{{} the
overall superposition state however. For an experimental value $\epsilon\sim0.42$
(where $\delta\sim0.4$), their separation is typically $\delta_{J}=0.4E/2$,
whereas the state $|\psi\rangle_{M}$ predicts a Gaussian distribution
for $J_{B}^{\theta_{B}}$, with $\Delta J_{B}^{\theta_{B}}\sim1.3E/2$.}

The $\delta$-scopic EPR nonlocality manifests without the significant
decoherence that normally prevents formation of Schrödinger cat states,
because the separation $\delta_{J}$ between states of the superposition
is not amplified relative to the quantum noise level. The EPR steering
parameter $\varepsilon_{\delta}$ is unchanged, consistent with the
requirement that entanglement cannot be created by local entangling
transformations, such as produced by beam splitters \cite{ent-measure}. 

\textbf{\emph{EPR nonlocality between distinct atom groups:}}\emph{
}The experiments of Refs. \cite{Plozik,Polzik2011EPR_macro_(e),muschik,muschik2}
investigate EPR entanglement between two spatially separated macroscopic
atomic ensembles, $A$ and $B$. In their experiment, $J_{A/B}^{X,Y,Z}$
are the collective spins of each ensemble, defined relative to two
atomic levels. The observation of the condition $D\equiv\{[\Delta(X_{A}-X_{B})]^{2}+\Delta[(P_{A}+P_{B})]^{2}\}/4<1$
implies entanglement between subsystems $A$ and $B$ \cite{duan}.
For spins, this entanglement condition becomes \cite{raymer}
\begin{eqnarray}
D\equiv\frac{[\Delta(J_{A}^{Z}+J_{B}^{Z})]+[\Delta(J_{A}^{Y}+J_{B}^{Y})]^{2}}{\bigl|\langle J_{A}^{X}\rangle\bigr|+\bigl|\langle J_{B}^{X}\rangle\bigr|} & < & 1.\label{eq:d}
\end{eqnarray}
Measurements give $D\sim0.8$ for thermal atomic ensembles \cite{Plozik,Polzik2011EPR_macro_(e)}.
The value of $D\sim0.5$ would imply an EPR steering nonlocality
(\ref{eq:schepr}) \cite{rrmp,beceprolsen,atoms-njp}. For a rigorous
demonstration of EPR nonlocality, it is however necessary to measure
the EPR observables independently and locally, so that information
is gained simultaneously about each of $J_{A}^{\theta_{A}}$ and $J_{B}^{\theta_{B}}$.

\begin{figure}[b]
\begin{centering}
\includegraphics[width=1\columnwidth]{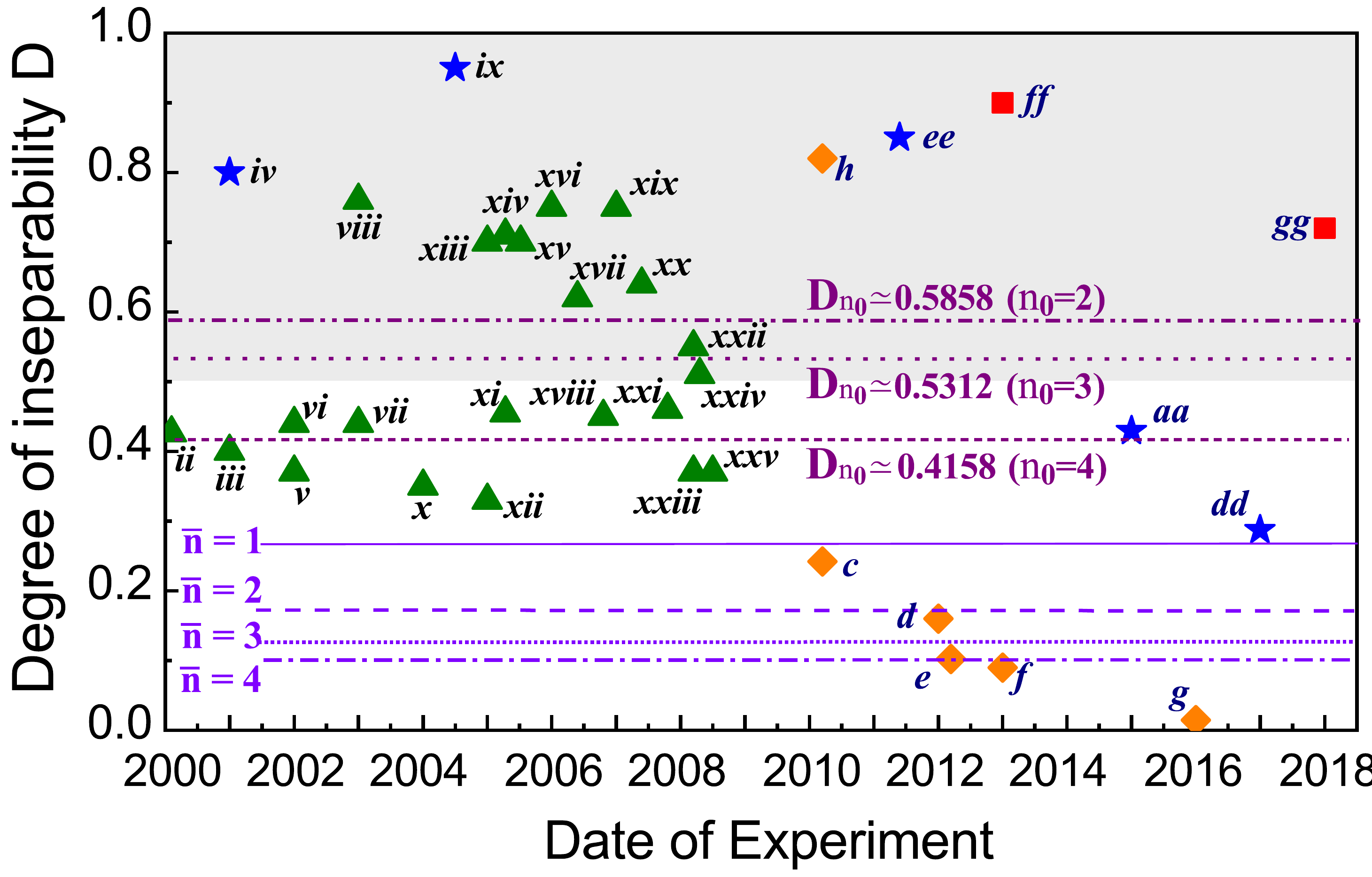}
\par\end{centering}
\caption{\textcolor{black}{\emph{ }}\textcolor{black}{The entanglement parameter
$D$ is plotted versus date for a sample of experiments.}\textcolor{black}{\emph{
}}\textcolor{black}{  Data ($ii-xxv$) are the values reported for
atomic (stars) and optics (triangles) experiments respectively, as
listed in Fig. 9 of Ref. \cite{rrmp}. Data $c$ (\cite{pengbest_(b)}),
$d$ (\cite{recentD_(c)}), $e$ (\cite{Hannover2012best_(d)}), $f$
(\cite{Hannover2013best}), $g$(\cite{photon-pair-2016}), $h$ (\cite{Aoki_2010_(a)})
are for optical experiments (diamonds), data $aa$ (\cite{eprnaturecommun}),
$dd$ (\cite{Optica2017}), $ee$ (\cite{Polzik2011EPR_macro_(e)})
are for atomic or hybrid experiments (stars), and $ff$ (\cite{polamaki}),
$gg$ (\cite{recnet-osci-2}) are for mechanical oscillator experiments
(squares).  $D<D_{\overline{n}}^{(l)}$ implies the EPR state $|\psi\rangle$
has a mean number of bosons greater than $\overline{n}$. The $D_{\bar{n}}^{(l)}$
are plotted for $\overline{n}=1,2,3,4$. $D<D_{n_{0}}$ requires states
of more than $n_{0}$ bosons. Plotted is $D_{n_{0}}$ for $n_{0}=2,3,4$.
For $n_{0}=10$, $D_{n_{0}}\simeq0.2228$.}}
\end{figure}

EPR steering correlations meeting the condition (\ref{eq:epr-cond})
have however recently been observed for matter-wave fields created
with Bose-Einstein condensates (BEC) \cite{eprnaturecommun,EntAtoms,SteerAt-obert,treu-matteo,heidelepr}.
In the experiments \cite{eprnaturecommun,heidelepr,SteerAt-obert},
twin atom-beam states are generated by a parametric interaction \cite{heidelepr}.
Atoms are created in pairs, one into each group $A$ and $B$ that
correspond to different spins. The atom field quadratures $X_{A/B}$,
$P_{A/B}$ are measured by an atomic homodyne method, where the local
oscillators are a different, larger group consisting of $E=\langle a_{LO}^{\dagger}a_{LO}\rangle\sim\langle b_{LO}^{\dagger}b_{LO}\rangle>10^{2}$
atoms \cite{heidelepr}. \textcolor{black}{Experiments of Piese
et al observe EPR correlations }(\ref{eq:epr-cond})\textcolor{black}{{}
between mesoscopic atomic groups $A$ and $B$ with $\varepsilon\sim0.85$
(with no spatial separation) \cite{eprnaturecommun}. Recently, Kunkel
et al observe an EPR steering with $\varepsilon\sim0.71$  between
atom groups spatially separated by $\sim10\mu m$ \cite{SteerAt-obert}.
Assuming results are unchanged if the experiment were reconfigured
along the lines of the Schwinger spin experiments, and that the distributions
are approximately gaussian, this suggests nonlocalities with $\delta_{J}>E\delta>20$
atoms. }

\textcolor{black}{Using a different method of generation, Fadel et
al observe EPR steering correlations ($\varepsilon\sim0.74$) between
the Schwinger spins $J_{A}^{Z}$, $J_{B}^{Z}$ (and $J_{A}^{Y}$,
$J_{B}^{Y}$) of two atomic groups separated by $\sim4$$\mu m$ \cite{treu-matteo}.
Here, group $A$ ($B$) consists of two BEC components, $a_{\pm}$
($b_{\pm}$). Their measurement of spins ($J^{X}$ or $J^{Y}$) is
achieved with a variable pulse rotation $\theta$, in analogy to the
polarising beam splitter of Bowen et al. (although without independent
selection of the two measurement angles). The BEC experiments thus
reveal quantifiable nonlocal indeterminacies in the }\textcolor{black}{\emph{atom}}\textcolor{black}{{}
number differences $J_{B}^{\theta_{B}}$ at the given site. The nonlocality
being tested here is whether an action at the site $A$ can create
a change in the number of atoms between the groups $b_{+}$ and $b_{-}$,
at site $B$.}

\textbf{\textcolor{black}{\emph{Quantification of number of bosons
in the two-mode entangled state: }}}\textcolor{black}{The large values
of $\delta_{J}$ arise for four-mode states. However, the }\textcolor{black}{\emph{two-mode}}\textcolor{black}{{}
EPR state $|\psi\rangle$ can itself be constrained to have a certain
degree of ``largeness''. For any}\textcolor{black}{\emph{ entangled}}\textcolor{black}{{}
 pure state $|\psi\rangle$, the mean total number of bosons is $\bar{n}=\langle\psi|a^{\dagger}a+b^{\dagger}b|\psi\rangle$.
}\textcolor{magenta}{}\textcolor{black}{{} The value of $D$ places
a lower bound on $\bar{n}$.}\textcolor{red}{{} }Using the identity
$|\langle ab\rangle|\leq\sqrt{(\langle a^{\dagger}a\rangle+1)\langle b^{\dagger}b\rangle}$
\cite{hillezub}, we find $D\geq D_{\overline{n}}^{(l)}$ \cite{supp},
where $D_{\bar{n}}^{(l)}=1+\bar{n}-\bar{n}\sqrt{1+2/\bar{n}}$ decreases
with $\bar{n}$, and is achieved for the two-mode squeezed state for
which $D=(1-x)/(1+x)$. In an experiment, the two-mode system is
generally not a pure state. The measured $\langle a^{\dagger}a+b^{\dagger}b\rangle$
does not then reflect the mean number of bosons in an entangled state,
because there may be components of the mixture that are not entangled.
However, the observation of $D<D_{\bar{n}}^{(l)}$ certifies that
a pure entangled state $|\psi\rangle$ with $\langle a^{\dagger}a+b^{\dagger}b\rangle>\bar{n}$
must be a component of the mixed state (refer \cite{supp}). Similarly,
by expanding all pure states in the basis of number states $|i\rangle_{a}|j\rangle_{b}$,
we prove in the Supplemental Materials that the value $D$ places
a lower bound on the minimum number of bosons $n_{0}=i+j$ contributing
a nonzero term to the expan\textcolor{black}{sion. If the bosons
are atoms, the state $|i\rangle_{a}|j\rangle_{b}$ has an entanglement
depth of $n_{0}=i+j$, meaning all $n_{0}$ atoms are mutually entangled
\cite{sorsolm-1,multi,supp}. }

\textcolor{black}{Experimental values of $D$ are plotted in Fig.
2. The values ${\color{red}{\color{black}D<0.2228}}$ confirm two-mode
optical EPR states $|\psi\rangle$ involving more than $10$ photons
($n_{0}>10$). These states are different to states constructed from
photon pairs, for which $n_{0}\leq2$. Where $D<0.5,$ the two modes
of the state $|\psi\rangle$ are both EPR steerable \cite{qi-discord-two-way-steer}.
Measurements by Piese et al \cite{eprnaturecommun} observe $D<0.43$,
implying}\textcolor{black}{\emph{ }}\textcolor{black}{two-way}\textcolor{black}{\emph{
}}\textcolor{black}{EPR steerable states $|\psi\rangle$ with more
than $3$ atoms (if spatial separation could be achieved) \cite{supp}.}

\textbf{\textcolor{black}{\emph{Conclusion:}}}\textcolor{black}{{} We
have given evidence for a mesoscopic EPR nonlocality that ``delocalises''
$\delta_{J}\sim10^{5}$ photons between two polarisation modes at
a given site. This represents a tenth of the full range of outcomes
(defined as that within $3$ standard deviations of the mean) for
the polarisation photon-number difference at the site. Recent experiments
with Bose-Einstein condensates show similar EPR nonlocalities involving
four atomic modes. This motivates new experiments where it may be
feasible to demonstrate an EPR nonlocality ``delocalising'' $\delta_{J}\sim10$
atoms across two highly-occupied atomic modes at a given site. The
criteria presented in this paper may also have a practical application.
The curves of Fig. 1 can be used to detect a genuine EPR effect, even
when a causal effect is present. If the maximum disturbance due to
the causal effect can be quantified (to be $\delta_{C}$ say), then
an EPR nonlocality can be deduced if $\epsilon_{\delta}<1$ where
$\delta>\delta_{C}$. An example of such a causal effect (``cross-talk'')
is given in Ref. \cite{treu-matteo}.  }
\begin{acknowledgments}
This research has been supported by the Australian Research Council
Discovery Project Grants schemes under Grant DP180102470. \textcolor{black}{M.D.R
thanks the hospitality of the Institute for Atomic and Molecular Physics
(ITAMP) at Harvard University.} Q.Y.H. thanks the National Key R\&D
Program of China (Grants No. 2018YFB1107200 and No. 2016YFA0301302)
and the National Natural Science Foundation of China (Grants No. 11622428
and No. 61675007).
\end{acknowledgments}


\begin{thebibliography}{References}
\bibitem{epr}A. Einstein, B. Podolsky, and N. Rosen, Phys. Rev.
\textbf{47}, 777 (1935).

\bibitem{Bell}J. S. Bell, Physics \textbf{1}, 195 (1964).

\bibitem{bell-ep}S. J. Freedman and J. F. Clauser,  Phys. Rev. Lett.
\textbf{28}, 938 (1972); A. Aspect, J. Dalibard, and G. Roger, Phys.
Rev. Lett. \textbf{49}, 1804 (1982); G. Weihs, T. Jennewein, C. Simon,
H. Weinfurter, and A. Zeilinger, Phys. Rev. Lett. \textbf{81}, 5039
(1998). 

\bibitem{loophole}\textcolor{black}{B. Hensen }\textcolor{black}{\emph{et
al}}\textcolor{black}{.,  Nature }\textbf{\textcolor{black}{526}}\textcolor{black}{,
682 (2015); M. Giustina,}\textcolor{black}{\emph{ }}\textcolor{black}{et
al., Phys. Rev. Lett. }\textbf{\textcolor{black}{115}}\textcolor{black}{,
250401 (2015); L. K. Shalm, et al., Phys. Rev. Lett. }\textbf{\textcolor{black}{115}}\textcolor{black}{,
250402 (2015); C. Abellán, W. Amaya, D. Mitrani, V. Pruneri, and M.
W. Mitchell, Phys. Rev. Lett. }\textbf{\textcolor{black}{115}}\textcolor{black}{,
250403 (2015).}

\bibitem{higher-n-bell}\foreignlanguage{australian}{N. D. Mermin,
Phys. Rev. D \textbf{22}, 356 (1980); J. C. Howell, A. Lamas-Linares,
and D. Bouwmeester, Phys. Rev. Lett. \textbf{88}, 030401 (2002); P.
D. Drummond, Phys. Rev. Lett. \textbf{50}, 1407 (1983); M. D. Reid,
W. J. Munro, and F. De Martini, Phys. Rev. A \textbf{66}, 033801 (2002);
D. Collins, N. Gisin, S. Popescu, D. Roberts, and V. Scarani,  Phys.
Rev. Lett. \textbf{88,} 170405 (2002); Q. Y. He, P. D. Drummond, and
M. D Reid, Phys. Rev. A \textbf{83}, 032120 (2011); R. F. Werner and
M. M. Wolf, Phys. Rev. A \textbf{64}, 032112 (2001); G. Tóth, O.
Gühne, and H. J. Briegel, Phys. Rev. A \textbf{73}, 022303 (2006);
O. Gühne, G. Tóth, P. Hyllus, and H. J. Briegel, Phys. Rev. Lett.
\textbf{95}, 120405 (2005); See also J. Tura, R. Augusiak, A. B. Sainz,
T. Vértesi, M. Lewenstein, A. Acín, Science\textbf{ 344}, 1256 (2014).}

\bibitem{cv-bell}\textcolor{black}{A different regime is provided
by continuous variable tests of LR. U. Leonhardt and J. Vaccaro, J.
Mod. Opt. }\textbf{\textcolor{black}{42}}\textcolor{black}{, 939 (1995).
A. Gilchrist, P. Deuar and M. Reid, Phys. Rev. Lett. }\textbf{\textcolor{black}{80}}\textcolor{black}{{}
3169 (1998). K. Banaszek and K. Wodkiewicz, Phys. Rev. Lett. }\textbf{\textcolor{black}{82}}\textcolor{black}{{}
2}009 (1999). A. Gilchrist, P. Deuar and M. D. Reid, Phys. Rev. A\textbf{60},
4259 (1999).\textcolor{red}{{} }C. Wildfeuer, A. Lund and J. Dowling,
Phys. Rev. A \textbf{76}, 052101 (2007). F. Toppel, M. V. Chekhova,
and G. Leuchs, arXiv:1607.01296 {[}quant-ph{]} (2016).\textcolor{red}{{}
}M. D. Reid,  Phys. Rev. A \textbf{97}, 042113 (2018). Oliver Thearle
et al.,  Phys. Rev. Lett. \textbf{120}, 040406 (2018).

\bibitem{schosteer}E. Schrödinger, Math. Proc. Camb. Phil. Soc.
\textbf{31}, 555 (1935).

\bibitem{hw}H. M. Wiseman, S. J. Jones, and A. C. Doherty, Phys.
Rev. Lett. \textbf{98}, 140402 (2007).

\bibitem{eprr}M. D. Reid, Phys. Rev. A \textbf{40}, 913 (1989).

\bibitem{steereric}E. G. Cavalcanti, S. J. Jones, H. M. Wiseman,
and M. D. Reid,  Phys. Rev. A\textbf{ 80}, 032112 (2009).

\bibitem{saunexp}D. J. Saunders, S. J. Jones, H. M. Wiseman, and
G. J. Pryde,  Nature Phys. \textbf{6}, 845 (2010). 

\bibitem{ou} Z. Y. Ou, S. F. Pereira, H. J. Kimble, and K. C. Peng,
Phys. Rev. Lett.\textbf{ 68}, 3663 (1992).

\bibitem{rrmp}M. D. Reid, P. D. Drummond, W. P. Bowen, E. G. Cavalcanti,
P. K. Lam, H. A. Bachor, U. L. Andersen, and G. Leuchs, Rev. Mod.
Phys. \textbf{81}, 1727 (2009),\textcolor{black}{{} and experiments
referenced therein.}

\bibitem{spinpolepr}W. P. Bowen, N. Treps, R. Schnabel, and P. K.
Lam, Phys. Rev. Lett. \textbf{89}, 253601 (2002); W. P. Bowen, R.
Schnabel, H.-A. Bachor, and P. K. Lam, Phys. Rev. Lett. \textbf{88},
093601 (2002).

\bibitem{Aiko_2010_PRA_exp(aa)} B. Hage, A. Samblowski, and R. Schnabel,
Phys. Rev. A \textbf{81}, 062301 (2010).\textcolor{blue}{{} }\textcolor{black}{}

\bibitem{Tobias2011_cc}\textcolor{black}{T. Eberle, V. Händchen,
J. Duhme, T. Franz, R. F. Werner, and R. Schnabel, Phys. Rev. A }\textbf{\textcolor{black}{83}}\textcolor{black}{,
052329 (2011). }

\bibitem{Walborn2011}\textcolor{black}{S. P. Walborn, A. Salles,
R. M. Gomes, F. Toscano, and P. H. Souto Ribeiro, Phys. Rev. Lett.
}\textbf{\textcolor{black}{106}}\textcolor{black}{, 130402 (2011).}

\bibitem{Aoki_2010_(a)}\textcolor{black}{A. Samblowski, C. E. Laukötter,
N. Grosse, P. K. Lam, and R. Schnabel,  arXiv:1011.5766v2, AIP Conference
Proceedings }\textbf{\textcolor{black}{1363}}\textcolor{black}{, 219
(2011).}

\bibitem{Hannover2012best_(d)}\textcolor{black}{S. Steinlechner,
J. Bauchrowitz, T. Eberle, and R. Schnabel, Phys. Rev. A }\textbf{\textcolor{black}{87}}\textcolor{black}{,
022104 (2013).}

\bibitem{Hannover2013best}\textcolor{black}{T. Eberle, V. Händchen,
and R. Schnabel,  Opt. Express }\textbf{\textcolor{black}{21}}\textcolor{black}{,
11546 (2013).}

\bibitem{Howell2013}\textcolor{black}{J. Schneeloch, P. B. Dixon,
G. A. Howland, C. J. Broadbent, and J. C. Howell,  Phys. Rev. Lett.}\textbf{\textcolor{black}{{}
110}}\textcolor{black}{, 130407 (2013).}

\bibitem{pengbest_(b)}\textcolor{black}{Y. Wang, H. Shen, X. Jin,
X. Su, C. Xie, and K. Peng, Opt. Express }\textbf{\textcolor{black}{18}}\textcolor{black}{,
6149 (2010).} \textcolor{black}{}

\bibitem{recentD_(c)}Z. Yan\textit{,} X. Jia, X. Su, Z. Duan, C.
Xie, and K. Peng,Phys. Rev. A \textbf{85}, 040305(R) (2012). \textcolor{black}{}

\bibitem{photon-pair-2016}J-C. Lee, K-K. Park, T-M. Zhao, and Y-H.
Kim,  Phys. Rev. Lett., \textbf{117}, 250501 (2016).\textcolor{red}{{}
}\textcolor{black}{}

\bibitem{eprnaturecommun}\textcolor{black}{J. Peise, I. Kruse, K.
Lange, B. Lücke, L. Pezzè, J. Arlt, W. Ertmer, K. Hammerer, L. Santos,
A. Smerzi, and C. Klempt,  Nat. Commun. }\textbf{\textcolor{black}{6}}\textcolor{black}{,
8984 (2015).}

\bibitem{SteerAt-obert}\textcolor{black}{P. Kunkel, M. Pr\"ufer,
H. Strobel, D. Linnemann, A. Fr\"olian, T. Gasenzer, M. G\"arttner,
and M. K. Oberthaler, }\textcolor{black}{{} Science }\textbf{\textcolor{black}{360}}\textcolor{black}{,
413 (2018). }

\bibitem{treu-matteo}\textcolor{black}{M. Fadel, T. Zibold, B. D\'ecamps,
and P. Treutlein, Science }\textbf{\textcolor{black}{360}}\textcolor{black}{,
409 (2018).}\textcolor{red}{{} }\textcolor{black}{}

\bibitem{Plozik}\textcolor{black}{B. Julsgaard, A. Kozhekin, and
E. S. Polzik, Nature }\textbf{\textcolor{black}{413}}\textcolor{black}{,
400 (2001).}

\bibitem{Polzik2011EPR_macro_(e)}\textcolor{black}{H. Krauter, C.
A. Muschik, Kasper Jensen, W. Wasilewski, J. M. Petersen, J. I. Cirac,
and E. S. Polzik, Phys. Rev. Lett. }\textbf{\textcolor{black}{107}}\textcolor{black}{,
080503 (2011).}

\bibitem{muschik}C. A. Muschik, E. S. Polzik, and J. I. Cirac, Phys.
Rev. A \textbf{83}, 052312 (2011).

\bibitem{muschik2}C. A. Muschik, H. Krauter, K. Jensen, J. M. Petersen,
J. I. Cirac, and E. S. Polzik,  J. Phys. B: At. Mol. Opt. Phys. \textbf{45},
124021 (2012).

\bibitem{Optica2017}\textcolor{black}{M. Dabrowski, M. Parniak, and
W. Wasilewski, Optica }\textbf{\textcolor{black}{4}}\textcolor{black}{,
272 (2017). }

\bibitem{paulo}V. Giovannetti, S. Mancini, and P. Tombesi, Europhys.
Lett.\textbf{\textit{ }}\textbf{54}, 559 (2001). 

\bibitem{eprentosc}S. G. Hofer,\textit{ }W. Wieczorek, M. Aspelmeyer,
and K. Hammerer,  Phys. Rev. A\textbf{ 84}, 052327 (2011).

\bibitem{polamaki}T. A. Palomaki, J. D. Teufel, R. W. Simmonds, and
K. W. Lehnert, Science \textbf{342}, 710 (2013).\textcolor{red}{{}
}\textcolor{black}{}

\bibitem{roman-schnabel-ent-epr-objects}R. Schnabel,  Phys. Rev.
A \textbf{92}, 012126 (2015).

\bibitem{recent-ent-osc}\textcolor{black}{R. Riedinger, A. Wallucks,
I. Marinkovi\'{c}, C. Löschnauer, M. Aspelmeyer, S. Hong, and S. Gröblacher,
Nature }\textbf{\textcolor{black}{556}}\textcolor{black}{, 473 (2018).
}\textcolor{red}{}

\bibitem{recnet-osci-2}\textcolor{black}{C. F. Ockeloen-Korppi, E.
Damskägg, J.-M. Pirkkalainen, M. Asjad, A. A. Clerk, F. Massel, M.
J. Woolley, and M. A. Sillanpää,  Nature }\textbf{\textcolor{black}{556}}\textcolor{black}{,
478 (2018).}\textcolor{red}{{} }\textcolor{black}{}

\bibitem{bell-osc}I. Marinkovi\'{c}, A. Wallucks, R. Riedinger, S.
Hong, M. Aspelmeyer, and S. Gröblacher, Phys. Rev. Lett. \textbf{121},
220404 (2018).

\bibitem{HeReid2013}\textcolor{black}{Q. Y. He and M. D. Reid, Phys.
Rev. A }\textbf{\textcolor{black}{88}}\textcolor{black}{, 052121 (2013). }

\bibitem{Simon2014}\textcolor{black}{S. Kiesewetter, Q. Y. He, P.
D. Drummond, and M. D. Reid,  Phys. Rev. A }\textbf{\textcolor{black}{90}}\textcolor{black}{,
043805 (2014).}

\bibitem{sorsolm-1}A. S. Sørensen and K. Mølmer, \textit{\emph{Phys.
Rev. Lett.}} \textbf{86}, 4431 (2001).

\bibitem{multi}C. Gross, T. Zibold, E. Nicklas, J. Esteve and M.
K. Oberthaler, Nature (London) \textbf{464}, 1165 (2010); M. F. Riedel,
P. Böhi, Y. Li, T. W. Hänsch, A. Sinatra and P. Treutlein,  Nature
(London) \textbf{464}, 1170 (2010). 

\bibitem{other-measures}A. Elitzer, S. Popescu and D. Rohrlich, Physics
Letters A \textbf{162}, 25 (1992). S. Portmann, C. Branciard, and
N. Gisin, Phys. Rev. A \textbf{86}, 012104 (2012).

\bibitem{Schrodinger}E. Schrödinger, Naturwissenschaften \textbf{23},
844 (1935).

\bibitem{LG}A. J. Leggett and A. Garg, Phys. Rev. Lett. \textbf{54},
857 (1985).

\bibitem{mlr}M. D. Reid, Phys. Rev. Lett. \textbf{84}, 2765 (2000);
M. Reid, in Proceedings of the 3rd Workshop on Mysteries, Puzzles
and Paradoxes in Quantum Mechanics, Gargnano, 2000, editors R. Bonifacio,
B. G. Englert and D. Vitali. p220; 

\bibitem{recent-mlr}M. D. Reid, Phys. Rev. A\textbf{ 97}, 042113
(2018).

\bibitem{cv-review}C. Weedbrook \emph{et al}., Rev. Mod. Phys.\textbf{
84}, 621 (2012).

\bibitem{gauscv}V. D\textquoteright Auria\textcolor{black}{, S. Fornaro,
A. Porzio, S. Solimeno, S. Olivares, and M. G. A. Paris,} Phys. Rev.
Lett. \textbf{102}, 020502 (2009).

\bibitem{supp}See Supplemental Material for details of all proofs.

\bibitem{cat-sup-in-cv}E. G. Cavalcanti and M. D. Reid, Phys. Rev.
Lett., \textbf{97}, 170405\textit{\emph{ (2006);}} C. Marquardt et
al., Phys. Rev. A \textbf{76} 030101 (2007); B. Opanchuk, L. Rosales-Zarate,
R. Y. Teh, and M. D. Reid, Phys. Rev. A \textbf{94,} 062125 (2016).

\bibitem{cavalr}E. G. Cavalcanti and M. D. Reid, Phys. Rev. A \textbf{77},
062108 (2008).

\bibitem{frowis}F. Fr{\"o}wis, P. Sekatski, D. Pavel and W. D{\"u}r,
Phys. Rev. Lett. \textbf{116} 090801 (2016). F. Fr{\"o}wis, N. Sangouard
and N. Gisin, Optics Communications \textbf{337}, 2 (2015).

\bibitem{flor-cat-rmp}F. Fröwis, P. Sekatski, W. Dür, N. Gisin, and
N.Sangouard, Rev. Mod. Phys. \textbf{90}, 025004 (2018).

\bibitem{probeplanck}I. Pikovski\textcolor{black}{,} M. R. Vanner,
M. Aspelmeyer, M. S. Kim, and \v{C}. Brukner, Nature Phys\textit{.}
\textbf{8}, 393 (2012).

\bibitem{ent-measure}V. Vedral\textcolor{black}{,} M. B. Plenio,
M. A. Rippin, and P. L. Knight,  Phys. Rev. Lett. \textbf{78}, 2275
(1997).

\bibitem{duan}L.-M.\emph{ }Duan\textit{, }G. Giedke, J. I. Cirac,
and P. Zoller, Phys. Rev. Lett. \textbf{84}, 2722 (2000).

\bibitem{raymer}M. G. Raymer, A. C. Funk, B. C. Sanders, and H. de
Guise,  Phys. Rev. A \textbf{67}, 052104 (2003).

\bibitem{atoms-njp}Q. Y. He and M. D. Reid, TNew J. Phys., \textbf{15},063027
(2013).

\bibitem{beceprolsen} A. J. Ferris\textit{,} M. K. Olsen, E. G. Cavalcanti,
and M. J. Davis, Phys. Rev. A \textbf{78}, 060104(R) (2008).

\bibitem{EntAtoms}\textcolor{black}{K. Lange, J. Peise, B. L\"ucke,
I. Kruse, G. Vitagliano, I. Apellaniz, M. Kleinmann, G. T\'oth, and
C. Klempt, Science }\textbf{\textcolor{black}{360}}\textcolor{black}{,
416 (2018).}\textcolor{red}{ }\textcolor{black}{}

\bibitem{heidelepr}C. Gross, H. Strobel, E. Nicklas, T. Zibold, N.
Bar-Gill, G. Kurizki, and M. K. Oberthaler, Nature \textbf{480},
219 (2011).

\bibitem{hillezub}M. Hillery and M. S. Zubairy, Phys. Rev. Lett.
\textbf{96}, 050503 (2006).

\bibitem{qi-discord-two-way-steer}Q. Y. He, Q. H. Gong and M. D.
Reid,  Phys. Rev. Lett. \textbf{114}, 060402 (2015).
\end{thebibliography}
\end{document}